\documentstyle[11pt,newpasp,twoside,epsfig]{article}
\makeatletter

\newenvironment{figurehere}
  {\def\@captype{figure}}
  {}
\makeatother

\def\edcomment#1{\iffalse\marginpar{\raggedright\sl#1\/}\else\relax\fi} 
\reversemarginpar 

\begin{document} 
\title{Recent Breakthroughs in Detecting Neutron Star Binaries in Globular
Clusters}

\author{Peter D. Edmonds} 
\affil{Harvard-Smithsonian Center for Astrophysics, 60 Garden St, 
Cambridge, MA 02138, USA; pedmonds@cfa.harvard.edu} 

\begin{abstract} 
Binary stars have long been considered to play a crucial role in globular
cluster evolution, and offer the advantages of studying systems at the
same, well-determined distances. However, early search attempts were
consistently thwarted by crowding (particularly in the optical) and initial
detections were limited to small numbers of low-mass X-ray binaries (LMXBs)
and a handful of other systems. This resolution hurdle has been
dramatically overcome by the superb spatial resolution and sensitivity of
{\em HST} and {\em Chandra} (supported by advances in radio observations),
enabling the detection in {\em individual} clusters of more than 10, and in
some cases more than 100, binaries. This review will focus on detections of
neutron star binaries, including recent optical identifications, the
exciting discoveries of multiple LMXBs in quiescence (with the potential to
constrain neutron star equations of state) and the detections of
millisecond pulsars (MSPs) in X-ray and optical images.

\end{abstract}

\section{Introduction} 

The study of globular cluster binaries is motivated by their impact on the
dynamical evolution of globular clusters (Hut et al. 1992). The stellar
densities near the centers of globular clusters can reach as high as 10$^6$
stars per cubic parsec, and interactions between single and binary stars
then become inevitable. These interactions can act as a heating source by
the conversion of binary binding energy into stellar kinetic energy when,
for example, low-mass stars are ejected after binary collisions, helping to
stall or prevent core collapse. Binaries offer the opportunity to study
the results of stellar interactions, and act as crucial tracers of the
white dwarf and neutron star populations in clusters.

This review will focus on observations of neutron star binaries (LMXBs and
MSPs) with an emphasis on recent work using {\em HST} and {\em Chandra}. A
separate observational paper in these proceedings by A. Cool will discuss
cataclysmic variables (CVs) and chromospherically active main sequence
binaries.

\section{Low-Mass X-ray Binaries}

Low-Mass X-ray Binaries contain low-mass stars filling their Roche lobes
and transferring material via an accretion disk to a neutron star. Twelve
X-ray bright ($L_X\sim10^{36}$ erg s$^{-1}$) LMXBs have been known in
globulars since 1981 (se e.g. Hertz \& Grindlay 1983). Many of these
binaries are transient and show X-ray bursts like the LMXBs found in the
galactic disk, but LMXBs are $\sim$200 times more common in globulars than
in the field, a sure sign that interactions play a vital role in their
formation.  These sources have been easily detected in X-rays, because of
their high luminosity and because usually only one, at most, is found in
each cluster (other sources are much fainter).  The same ease of detection
does not apply in the optical where crowding near the centers of globulars
is a formidable problem, but optical identifications of LMXBs offer crucial
advantages over X-ray observations. They may, for example, allow direct
observation of the secondary star, and they often result in period
determinations, or confirmations of X-ray periods, crucial information for
understanding these systems and their evolution.

\begin{figure}
\hspace*{1.3cm}
\epsfig{file=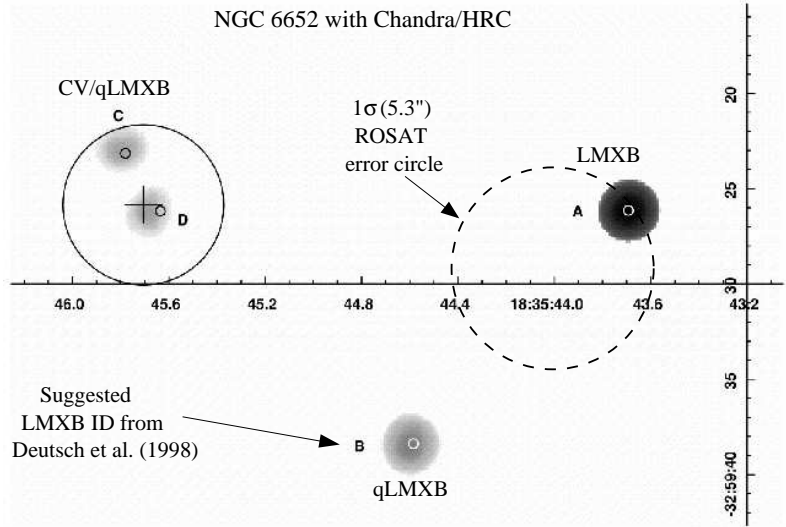,width=10.5cm}
\caption{{\em Chandra}/HRC image of NGC 6652 from Heinke et
al. (2001). Circles show blue, variable optical counterparts for three of
the 4 detected X-ray sources (A, B \& C). Variability in the labeled star
near D is not seen and this is not likely to be an optical ID.  The
5$\sigma$ ROSAT error circle is also shown.}
\end{figure}

The first optical counterpart to a globular cluster LMXB was the subgiant
AC211 discovered in M15 (Auriere et al. 1984; Ilovaisky et al. 1987). These
ground-based observations succeeded because AC211 is very bright optically,
and arguably no other secure LMXB optical counterpart was found before the
availability of {\em HST}. A comparison of ground-based M15 images (Auriere
et al. 1984) with an {\em HST} image (White \& Angelini 2001) shows the
benefits of using {\em HST} near the centers of dense globulars. Another
recent advance has been greatly improved LMXB X-ray positions thanks to the
subarcsecond resolution of {\em Chandra}. With multiple sources the
combination of {\em HST} and {\em Chandra} gives astrometry at the
0\farcs05 to 0\farcs2 level. Successful applications include the
identification by White \& Angelini (2001) of an optical counterpart for a
second LMXB in M15 and the identification by Heinke, Edmonds \& Grindlay
(2001) of optical IDs for 3 X-ray sources in NGC 6652 (see Fig. 1)
including the previously known LMXB, a likely LMXB in quiescence and a
third source (a qLMXB or CV). The qLMXB counterpart was previously
suggested by Deutsch, Margon \& Anderson (1998) and Deutsch, Margon \&
Anderson (2000) as an ID for the LMXB.

Eight optical IDs to LMXBs in 7 globulars have now been found: 2 in M15 and
1 each in NGC 6712, NGC 6624, NGC 1851, NGC 6441, NGC 6440 and NGC 6652
(see Homer et al. 2002 and references therein; {\em HST} was used in 6 of the
discovery or confirming papers, and {\em Chandra} in 5 of them).  With the
exception of NGC 6440 the E($B-V$) values are all $<0.5$ for these
clusters, while the clusters lacking LMXB optical IDs all have E($B-V$)
$>1.4$. Four of the 7 systems with period determinations have periods less
than 1 hr, while $< 10$\% of field LMXBs have periods this small (Deutsch
et al. 2000), another sign that interactions form these binaries.

The X-ray bright LMXBs represent only a small fraction of the total X-ray
source population in globular clusters. Based on Einstein observations,
Hertz \& Grindlay (1983) pointed out the bimodal nature of the cluster
X-ray luminosity function: the bright sources (accreting LMXBs) have $L_X >
10^{36}$ erg s$^{-1}$ and a population of low luminosity sources have $L_X
\sim 10^{32}-10^{34}$ erg s$^{-1}$.  Early hints that some of these low
luminosity X-ray sources are qLMXBs came from the observation that several
systems, such as the one in NGC 6440, are transients alternating between
high and low luminosity states (Hertz \& Grindlay 1983).  With the use of
{\em Chandra}, large numbers of these sources been finally been
unambiguously identified as a mixture of qLMXBs, CVs, MSPs and
chromospherically active binaries (Grindlay et al. 2001a;
hereafter GHE01a and Grindlay et al. 2001b; hereafter GHE01b).

Consider the case of the rich globular cluster 47~Tucanae. The five low
luminosity ROSAT sources detected near the center of this cluster are
heavily overlapped but with {\em Chandra} the bright sources are easily
resolved and dozens of fainter sources are detected. Figure 2 shows a
comparison between the {\em Chandra} image of GHE01a and the ROSAT contour
plot of Verbunt \& Hasinger (1998). Also shown is an X-ray `color magnitude
diagram' for this field, exploiting the spectral resolution and sensitivity
of {\em Chandra}. The two bright, relatively soft sources (X5 and X7) have
spectra consistent with neutron star atmospheres, as expected for qLMXBs,
and the bright, hard sources have spectra consistent with CVs
(GHE01a). Similar results have been obtained with the detection of one
qLMXB in the nearby, core-collapsed cluster NGC~6397 (GHE01b), and 4 or 5
qLMXBs in NGC 6440, a number broadly consistent with the high inferred
collision rate in this globular (Pooley et al. 2002).

The spectra of these qLMXBs are well fit by hydrogen atmosphere models of
neutron stars (plus in some cases weak power laws). This work has the
exciting capability to constrain the mass and radius, and hence the
equation of state, of the neutron stars in these binaries. Examples of
early work in this field include those of Rutledge et al. (2002) for the
qLMXB in $\omega$ Cen, GHE01b for the qLMXB in NGC~6397 and in't Zand et
al. (2001) for the transient LMXB in NGC 6440. The strongest constraints
have come from a detailed study of X5 and X7 in 47~Tuc, by Heinke et
al. (2002). These two qLMXBs are particularly interesting because they are
the brightest known qLMXBs in globulars (in terms of flux), allowing for
the most efficient observational constraints.

\begin{figurehere}
\hspace*{0.7cm}
\vspace*{-2.0cm}
\epsfig{file=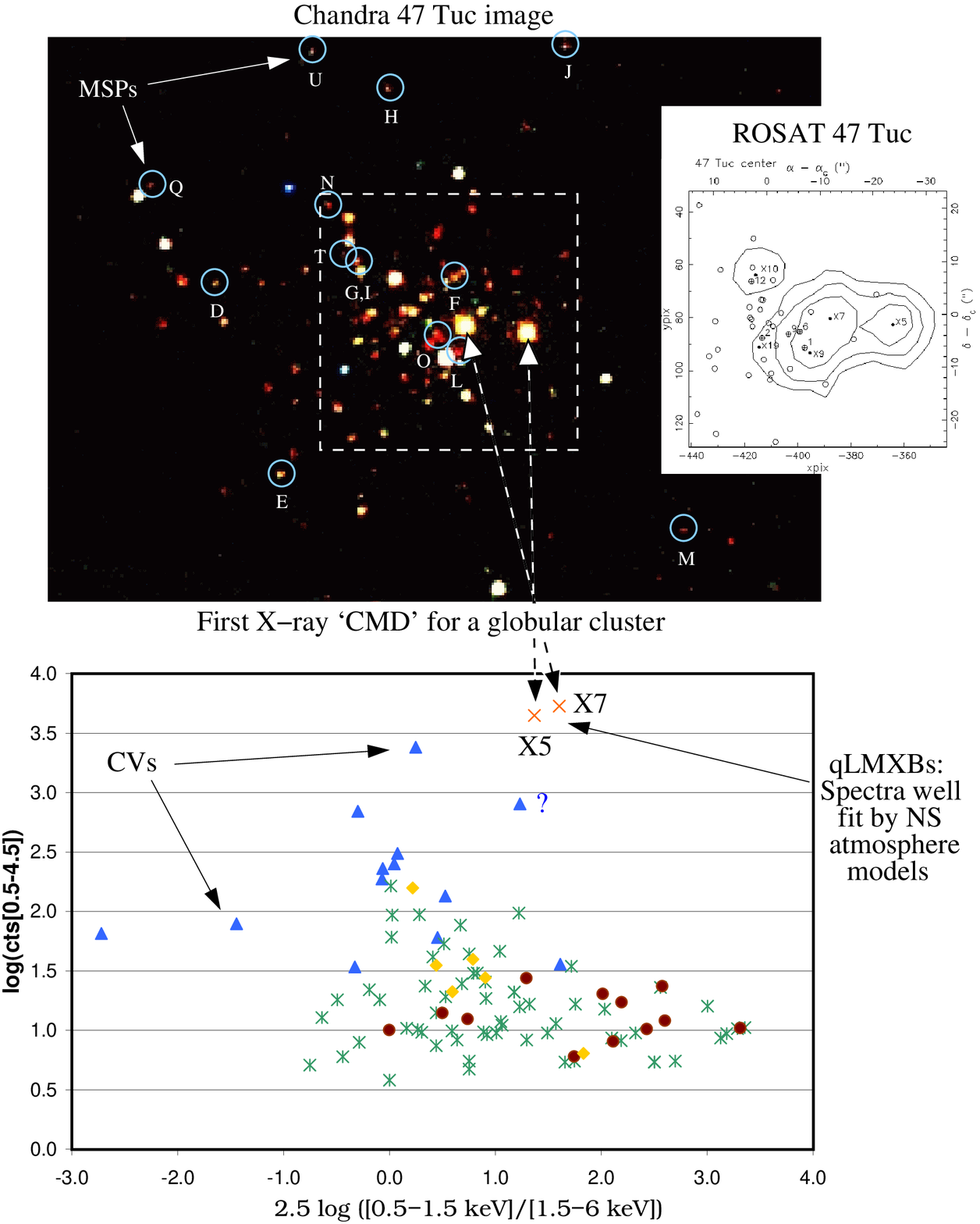,width=11.5cm}
\vspace*{2.0cm}
\caption{{\em Chandra} 47~Tuc image and X-ray color magnitude diagram from
GHE01a, with the qLMXBs (X5 and X7) and the MSPs from Camilo et al. (2000)
labeled. Also shown is a contour plot from the ROSAT observations of
Verbunt \& Hasinger (1998).}
\vspace*{1.0cm}
\end{figurehere}

With the exception of the transient in NGC 6440, none of these qLMXBs have
ever been detected in outburst. Because of this, the detection of optical
counterparts to these quiescent sources has been very challenging, and only
one successful ID has been found (X5; Edmonds et al. 2002a). The
counterpart to this eclipsing system (see Heinke et al. 2002 for X-ray
light curves) was identified using precise astrometry between {\em HST} and
{\em Chandra} and the discovery of a faint, blue variable star within the
X-ray and radio error circles (Edmonds et al. 2002a). Eventually, radial
velocity measurements of the X5 companion may lead to measurements of the
neutron star mass, however because the ID is faint and crowded it will be a
difficult target for either {\em HST} or ground-based telescopes. No
optical counterparts have been detected for either X7 in 47~Tuc (Edmonds
qet al. 2002a) or the qLMXB in NGC~6397 (GHE01b) despite the use of deep
{\em HST} imaging.

\section{Millisecond pulsars}

The canonical theory predicts that LMXBs evolve into MSPs, so if LMXBs are
over-abundant in globulars, MSPs should be as well. The first known
globular cluster MSP was discovered in M28 (Hamilton et al. 1985; Lyne et
al. 1987) using timing observations in the radio, and most detections of
MSPs in globular clusters have used the same technique, including the
discovery of 8 MSPs in M15 (Anderson 1993) and 11 in 47~Tuc (Manchester et
al. 1991; Robinson et al. 1995). Many of these MSPs are binaries, and since
they have followed a different formation path from field MSP binaries, are
interesting for evolutionary studies.  They may also act as a probe of MSP
irradiation and ablation, and with the detection of optical companions,
neutron star mass measurements are possible.

Recent developments in this rapidly evolving field include the detection of
many new MSPs with Parkes, multiple X-ray detections with {\em Chandra} and
the detections of several X-ray counterparts using {\em HST}. Key studies
of 47~Tuc have been performed by Camilo et al. (2000) and Freire et
al. (2001), almost doubling the number of detected MSPs in this cluster. Of
the 20 MSPs published by these authors, 13 are in binary systems and 10 of
these have good constraints on the secondary masses. These systems divide
evenly into 2 groups, the first containing systems with relatively long
periods of 0.4--2.3 days and relatively massive secondaries
($\sim0.2M_\odot$ ) and the second containing systems with short periods of
0.06--0.2 days and lower mass secondaries of $\sim0.03M_\odot$ (Camilo et
al 2000 and Freire et al. 2001). The first group are reasonably well
understood as being helium white dwarfs (confirmed optically in one case;
see below), but the second are not well understood, and may be low-mass
degenerate stars (Rasio, Pfahl \& Rappaport 2000) or perhaps even brown
dwarfs (Bildsten \& Chakrabarty 2001).

A crucial result of the Freire et al. (2001) study is the determination of
timing positions, accurate to a few milli-arcsec, for 15 of the 20
MSPs. These accurate positions are crucial in searching for counterparts to
the radio MSPs in X-ray and optical images (until recently only one
globular cluster MSP had been observed in X-rays, the pulsar in M28;
Danner, Kulkarni \& Thorsett 1994; Becker \& Tr{\" u}mper 1997). With the help
of {\em Chandra}'s exceptional sensitivity and spatial resolution, large
numbers of cluster MSPs have now been detected in X-rays, including most of
the MSPs in 47~Tuc (GHE01a and Grindlay et al. 2002; see Fig. 2). The X-ray
counterparts in this cluster are mostly soft sources, and thermal emission
from the MSP polar caps is thought to be the dominant emission mechanism
(Grindlay et al. 2002). X-ray counterparts to globular cluster MSPs have
also been found in NGC~6397 (GHE01b) and NGC 6752 (D'Amico et al. 2002).

Since many of the 47~Tuc MSPs have been detected both in the radio and in
X-rays, and many of the X-ray sources are CVs or active binaries visible in
{\em HST} images, the X-ray, optical and radio data can be placed on a
common astrometric system good to $
 \sim0\farcs1$. This astrometry
resulted in the detection, by Edmonds et al. (2001) of the first optical
counterpart to a globular cluster MSP, 47~Tuc~U, a binary pulsar with a
relatively high mass companion (thought to be a He WD) and a period of 0.43
days. With the use of the excellent astrometry available for 47~Tuc, a blue
star (having broadband colors consistent with theoretical He WD models) was
discovered as the obvious optical counterpart to 47~Tuc~U. Low amplitude
variability detected in this star has a period consistent with the radio
period of 47~Tuc~U (Edmonds et al. 2001).

The second detection of an optical companion to a globular MSP was made in
NGC~6397.  The only known MSP in this cluster (the 1.35 day period,
eclipsing PSR J1740-53 or 6397-A; D'Amico et al. 2001) has a radio position
coinciding with the position of a relatively bright, red variable star
(Ferraro et al. 2001). Assuming an ellipsoidal model for the variations of
this star, the optical variability agrees beautifully with the radio binary
period and phase, confirming the star as the optical companion to the MSP
(Ferraro et al. 2001). This star lies $\sim0.5$ mag below the subgiant
branch slightly redwards of the main sequence turnoff, and because it is
much brighter and less crowded than either X5 or the 47~Tuc~U companions it
has enormous potential for radial velocity work. It has already been the
subject of two detailed studies (Kaluzny, Rucinski \& Thompson 2002 and
Orosz \& van Kerkwijk 2002). A detection of 6397-A has also been made with
{\em Chandra} (GHE01b and Grindlay et al. 2002) including the discovery of
likely orbital modulation in the X-ray light curve.

Recently, a third detection of an optical companion to an MSP in a globular
cluster has been made using {\em HST} imaging. A faint star with large
amplitude variability and a period of $\sim$3.2 hrs was discovered as the
optical counterpart to the X-ray source W29 in 47~Tuc (Edmonds et
al. 2002b). Extensive time series analysis by these authors using the 8.3
day, almost continuous $V$ and $I$ time series of Gilliland et al. (2000),
plus archival data, have allowed the period and phase of this star
($\mathrm{W29_{opt}}$) to be measured to high levels of accuracy and these
orbital parameters were compared with those of the binary MSPs in Camilo et
al. (2000). The period of $\mathrm{W29_{opt}}$ was found to differ from
that of the eclipsing MSP 47~Tuc~W by only ($0.5\pm3.6$) s (!), a
percentage difference of 0.0045\%. Phase differences for the expected times
of optical maximum of only ($0.0008\pm0.0012$) days were found (Edmonds et
al. 2002b). This remarkable period and phase match between
$\mathrm{W29_{opt}}$ and 47~Tuc~W confirms $\mathrm{W29_{opt}}$ as the
optical companion without having a timing position for the MSP. These
orbital modulations are probably caused by MSP irradiation of one side of
the tidally locked companion.  This MSP companion is likely to be a main
sequence star, based on the optical photometry for this object and the
detection of radio eclipses by Camilo et al. (2000), and is likely to have
had an interesting dynamical history (Edmonds et al. 2002b).

\section{Summary and prospects}

Recently there has been a burst of activity in studies of neutron star
binaries in globular clusters, thanks mainly to the exquisite sensitivity
and spatial resolution of {\em HST} and {\em Chandra}.  These studies
include the detection of optical counterparts to most of the bright cluster
LMXBs, the identification of large numbers of qLMXBs (with exciting
potential for constraining neutron star equations of state), the discovery
of large numbers of MSPs in X-rays and the detection of optical
counterparts to three globular MSPs. 

Additional discoveries with deeper {\em Chandra} observations are sure to
be made, and e.g. the recent 300 ks 47~Tuc observation of J.~Grindlay and
collaborators will provide much higher signal-to-noise images and spectra
for the qLMXBs and MSPs. Much observational work remains to be done in
obtaining optical spectra of neutron star secondaries, including radial
velocity work. On the theoretical side, modeling of individual systems is
needed, such as that carried out already for 6397-A by Burderi et
al. (2002). Other theoretical goals include comparisons between clusters
(see the papers by A.~Cool and F.~Verbunt in these proceedings) and the
`ecological' modeling of entire globular clusters by self-consistently
including stellar and binary evolution and cluster dynamics (Portegies
Zwart et al. 1997).

\acknowledgments The author warmly acknowledges discussions with
collaborators including R.~Gilliland, F.~Camilo, H.~Cohn, A.~Cool,
J.~Grindlay, C.~Heinke \& P.~Lugger.

\end{document}